\documentclass[pre,endfloats]{revtex4}

\usepackage{amsfonts}
\usepackage{amsmath}
\usepackage{amssymb}
\usepackage{graphicx}
\newcommand{\apriori}{\textit{a priori}}

\newcommand{\etc}{etc.} 
\newcommand{\eg}{e.g.}
\newcommand{\RnD}{R\&D}

\newcommand{\mathperiod}{\quad .}
\newcommand{\mathcomma}{\quad ,}

\newcommand{\wraprefprepost}[3]{\wrapprepost{#1}{#2}{\ref{#3}}}
\newcommand{\formatrefplain}{\ref}
\newcommand{\formatrefparens}{\wraprefprepost{(}{)}}

\newcommand{\wrapprepost}[3]{{#1}{#3}{#2}}

\newcommand{\tagwithlabel}[2]{#1~#2}

\newcommand{\makelabeledcrossrefmacro}[4]
	{\newcommand{#3}{#1{#4}{#2}}}
\newcommand{\makecrossrefmaker}[3]
	{\newcommand{#1}{\makelabeledcrossrefmacro{#2}{#3}}}

\newcommand{\eqnrefformat}{\formatrefparens}
\newcommand{\eqnlabelbinding}{\tagwithlabel}
\newcommand{\eqnlabel}{eq.}
\newcommand{\Eqnlabel}{Eq.}
\newcommand{\eqnslabel}{eqs.}
\newcommand{\Eqnslabel}{Eqs.}

\newcommand{\eqnnum}{\eqnrefformat}
\makecrossrefmaker{\newlabeledeqnref}{\eqnlabelbinding}{\eqnnum}
\makecrossrefmaker{\newwordpluseqnref}{\tagwithlabel}{\eqnnum}

\newlabeledeqnref{\eqn}{\eqnlabel}
\newlabeledeqnref{\Eqn}{\Eqnlabel}
\newlabeledeqnref{\eqns}{\eqnslabel}
\newlabeledeqnref{\Eqns}{\Eqnslabel}

\newwordpluseqnref{\andeqn}{and}
\newwordpluseqnref{\througheqn}{through}

\newcommand{\figrefformat}{\formatrefplain}
\newcommand{\figlabelbinding}{\tagwithlabel}
\newcommand{\figlabel}{fig.}
\newcommand{\Figlabel}{Fig.}
\newcommand{\figslabel}{figs.}
\newcommand{\Figslabel}{Figs.}

\newcommand{\fignum}{\figrefformat}
\makecrossrefmaker{\newlabeledfigref}{\figlabelbinding}{\fignum}
\makecrossrefmaker{\newwordplusfigref}{\tagwithlabel}{\fignum}

\newlabeledfigref{\fig}{\figlabel}
\newlabeledfigref{\Fig}{\Figlabel}
\newlabeledfigref{\figs}{\figslabel}
\newlabeledfigref{\Figs}{\Figslabel}

\newwordplusfigref{\andfig}{and}
\newwordplusfigref{\throughfig}{through}

\newcommand{\sxnrefformat}{\formatrefplain}
\newcommand{\sxnlabelbinding}{\tagwithlabel}
\newcommand{\sxnlabel}{section}
\newcommand{\Sxnlabel}{Section}
\newcommand{\sxnslabel}{sections}
\newcommand{\Sxnslabel}{Sections}

\newcommand{\sxnnum}{\sxnrefformat}
\makecrossrefmaker{\newlabeledsxnref}{\sxnlabelbinding}{\sxnnum}
\makecrossrefmaker{\newwordplussxnref}{\tagwithlabel}{\sxnnum}

\newlabeledsxnref{\sxn}{\sxnlabel}
\newlabeledsxnref{\Sxn}{\Sxnlabel}
\newlabeledsxnref{\sxns}{\sxnslabel}
\newlabeledsxnref{\Sxns}{\Sxnslabel}
	
\newwordplussxnref{\andsxn}{and}
\newwordplussxnref{\throughsxn}{through}

\newcommand{\tblrefformat}{\formatrefplain}
\newcommand{\tbllabelbinding}{\tagwithlabel}
\newcommand{\tbllabel}{table}
\newcommand{\Tbllabel}{Table}
\newcommand{\tblslabel}{tables}
\newcommand{\Tblslabel}{Tables}

\newcommand{\tblnum}{\tblrefformat}
\makecrossrefmaker{\newlabeledtblref}{\tbllabelbinding}{\tblnum}
\makecrossrefmaker{\newwordplustblref}{\tagwithlabel}{\tblnum}

\newlabeledtblref{\tbl}{\tbllabel}
\newlabeledtblref{\Tbl}{\Tbllabel}
\newlabeledtblref{\tbls}{\tblslabel}
\newlabeledtblref{\Tbls}{\Tblslabel}
	
\newwordplustblref{\andtbl}{and}
\newwordplustblref{\throughtbl}{through}

\newcommand{\figpath}{}

\begin{document}

\title{The Network of EU-Funded Collaborative R\&D Projects }

\author{M. J. Barber}
\affiliation{Centro de Ci\^encias Matem\'aticas, Universidade da Madeira, Funchal, Portugal}
\author{A. Krueger}
\author{T. Krueger}
\affiliation{Universit\"at Bielefeld, Bielefeld, Germany}
\author{T. Roediger-Schluga}
\affiliation{ARC systems research, Vienna, Austria}
%\keywords{one two three}
%\pacs{PACS number}

\begin{abstract}
We describe collaboration networks consisting of research projects funded by the 
European Union and the organizations involved in those projects. The networks are 
of substantial size and complexity, but are important to understand due to the 
significant impact they could have on research policies and national economies in the EU.
In empirical determinations of the network properties, we observe characteristics similar 
to other collaboration networks, including scale-free degree distributions, small diameter, 
and high clustering. We present some plausible models for the formation and structure of
networks with the observed properties. 
\end{abstract}

\volumeyear{year}
\volumenumber{number}
\issuenumber{number}
\eid{identifier}

\date{\today}

\maketitle

%\tableofcontents

\section{Introduction}

Real world network analysis has become a major issue of research in the last
years. Most prominent are perhaps the investigations of the structure of the
World Wide Web, the network of internet routers, and certain social networks like citation
networks. On the theoretical side, one tries to understand the mechanisms of
formation of such networks and to derive statistical properties of the networks 
from the
generating rules. On the rigorous mathematical side, there are only a few
results for specific models, indicating the difficulty of a purely mathematical
approach (for a survey of recent results in this direction, see \cite{5}).
Thus, the main approach is to use some mean field assumption to get relevant
information about the corresponding graphs. Although it is not clear where the
limits of this approach lie, in many cases the results match well with
numerical simulations and empirical data. 

In this article, we study a particular collaboration network. Its vertices
are research projects funded by the European Union and the
organizations involved in those projects. In total, the data base contains over 
20000 projects
and 35000 participating organizations. The network shows all the
main characteristics known from other complex network structures,
such as scale-free degree distribution, small diameter, high clustering, and
inhomogeneous vertex correlations. 

Besides the general interest in studying a
new, real-world network of large size and high complexity, the study could have a significant economic impact. Improving collaboration between actors involved in innovation processes is a key objective of current science, technology, and innovation policy in industrialized countries. However, very little is known about what kind of network structures emerge from such initiatives. Moreover, it is quite likely that network structure affects network functions such as knowledge creation, knowledge diffusion, and the collaboration of particular types of actors. Presumably, this is determined by both endogenous formation mechanisms and exogenous framework conditions. In order to progress in our understanding, it is therefore essential to have sound statistics on the structure of networks we observe and to develop plausible models of how these are formed and evolve over time.

The model networks we use to compare
with the empirical data are random intersection graphs, a natural framework
for describing projections of bipartite graphs. Discrete intersection graphs
similar to the ones we use were first discussed in \cite{11}. We extend and
refine the construction from \cite{11} to be more applicable to real world graphs.

Perhaps the most important finding from our
model approach is the strong determination of the real network structure by
the degree distribution. That is, most statistical properties we measure in
the EU research project networks are the ones observed in a typical
realization of a uniform weighted random graph model with given (bipartite)
degree distribution as in the EU networks. Since this distribution is
characterized by two exponents---one for each partition---we have essentially
only four parameters (size, edge number, and exponents) which are needed to
describe the entire network. This is a tremendous reduction of complexity
indicating that only a few basic formation rules are driving the network
evolution. 

In \sxn{sec:dataset}, we describe the preparation of the data on the 
EU research programs. We present empirical determination of the network properties
in \sxn{sec:networkstructures}, followed by an explanation of these properties using 
a random intersection graph model in \sxn{sec:randgraphmodel}. Finally, 
in \sxn{sec:conclusions}, we summarize the key results and consider implications of the 
network properties on EU research programs.

\section{The data set} \label{sec:dataset}

In this work, we study research collaboration networks that have emerged in the European Union's first four successive four-year Framework Programs (FPs) on Research and Technological Development. Since their inception in 1984, six FPs have been launched, on four of which we have comprehensive data. FPs are organized in priority areas, which include information and communication technologies (ICTs), energy, industrial technologies, life sciences, environment, transportation, and a number of additional activities. In line with economic structural change, the main thematic focus of the FPs has shifted somewhat over time from energy and industrial technologies to the application of ICTs and life sciences. The majority of funding activities are aimed at stimulating research partnerships between firms, universities, research organizations, governmental actors, NGOs, lobby groups, \etc. Since FP4, the scope of activities has been expanded to also cover training, networking, demonstration, and preparatory activities (for details, see reference \cite{12}). In order to keep our data set compatible over the different FPs, we have excluded the latter set of projects from FP4 and only focus on collaborative research projects (see \tbl{tbl:frameworkprograms}).

In order to receive funding, projects in FP1 to FP4 had to comprise at least two organizations from at least two member states. We have retrieved data on these projects from the publicly available CORDIS (Community Research and Development Information Service) projects database \cite{2}. This database contains information on all funded projects as well as a reasonably complete listing of all participating organizations. 

The raw data on participating organizations is rather inconsistent. Apart from incoherent spelling in up to four languages per country, organizations are labelled inhomogeneously. Entries may range from large corporate groupings, such as Siemens, or large public research organizations like the Spanish CSIC to individual departments or labs and are listed as valid at the time the respective project was carried out. Among heterogeneous organizations, only a subset contains information on the unit actually participating or on geographical location (address, city, region and/or country). Information on older entries and the substructure of firms tends to be less complete.

Because of these difficulties, any automatic standardization method akin to the one utilized by Newman \cite{1} is inappropriate to this kind of data. Rather, the raw data has to be cleaned and completed manually, which is an ongoing project at ARC systems research. The objective of this work is to produce a data set useful for policy advice by identifying homogeneous, economically meaningful organizational entities. To this end, organizational boundaries are defined by legal control and entries are assigned to the respective organizations. Resulting heterogeneous organizations, such as universities, large research centres, or conglomerate firms are broken down into subentities that operate in fairly coherent areas of activity, such as faculties, institutes, divisions or subsidiaries. These can be identified for a large number of entries, based on the available contact information of participants, and are comparable across organizations.

The case of the French Centre National de la Recherche Scientifique (CNRS), the most active participant in the EU FPs may serve as an illustration. First, 785 separate entries were summarized under a unique organizational label. Next, these 785 entries were broken down into the eight areas of research activity in which CNRS is currently organized. Based on available information on participating units and geographical location, 732 of the 785 entries could be assigned to one of these subentities. For the remaining 53 entries, the nonspecific label CNRS was used.

Comparable success rates were achieved for other large public research organizations and universities. Due to scarcer information, firms could not be broken down at a comparable rate. Moreover, due to resource constraints, standardization work has focused on the major players in the FPs. Organizations participating in fewer than a total of 30 projects in FP1--4 have not been broken down yet. Due to these limitations in processing the data, we cannot rule out the possibility of a bias in analysing our data. However, we have run all the reported analyses with the undivided organizations and have obtained qualitatively similar results, apart from different extreme values, \eg, maximum degree.

\Tbl{tbl:frameworkprograms} displays information on the present data set, which contains information on a total of 27,758 projects, carried out over the period 1984 to 2004. It shows that the total budget as well as number of funded projects has increased dramatically from FP1 to FP4. Moreover, it provides a rough measure on the completeness of the available data. For a sizeable number of projects, the CORDIS project database lists information only on the project co-ordinator. This is due to the age of the data and inhomogeneous disclosure policies of different units at the European Commission. Comparing the number of projects containing information on more than one participant with the total number of projects funded in each FP shows that the data is fairly complete as of FP2.

The fact that FP1 was the first program launched and that the available data are rather incomplete make it exceptional in many respects. We therefore focus our analyses on FP2--4 and only give graph characteristic values for FP1 to indicate the difference to the networks created by the subsequent FPs.

\begin{table}[tbp] \centering
\begin{tabular}
	[c]{|c|c|c|c|c|c|c|}\hline
	\textbf{Framework Program } & 
	\textbf{budget\footnote{billion ECU/EUR}} &
	\textbf{\# P} & 
	\textbf{million EUR/P} & 
	\textbf{\#(P $>$1)\footnote{projects with more then one participating organization}} &
	\textbf{\# O} & 
	\textbf{million EUR/O}\\\hline
FP1 (1984--1988) & 3.8 & 3283 & 1.15 & 1696 & 2500 & 1.52\\
FP2 (1987--1991) & 5.4 & 3885 & 1.39 & 3013 & 6135 & 0.88\\
FP3 (1990--1994) & 6.65 & 5294 & 1.25 & 4611 & 9615 & 0.69\\
FP4\footnote{\RnD\ projects listed in parentheses. The number excludes all projects 
devoted to preparatory, demonstration, and training activities.} (1994--1998) & 13.3 &
15061 (9087) & 0.88 & 11374 (8039) & 20873 & 0.64\\\hline
\end{tabular} 
\caption{FP1--4 total budget and number of funded projects. The smaller average funding per project and org in FP4 is an artefact as it involves a large number of scholarships and the like, which are smaller than research projects (however, we cannot isolate the bias created).}
\label{tbl:frameworkprograms}
\end{table}%

\section{The network structure} \label{sec:networkstructures}

In this section, we present the basic properties of the network structure for
projects and organizations in the first four EU Framework Programs. We
consider both graphs as intersection graphs, each being the dual of the other,
which, for our purposes, is generally more convenient than the usual
bipartite-graph point of view. Recall that an intersection graph is given by
an enumerated collection of sets---the vertices of the intersection
graph---with elements from a given fixed base-set and edges defined via the
intersection property (edge $\triangleq$ nonempty intersection of two sets).
The sets need not be distinct. 

We denote by $\mathcal{P}=\left\{  P_{1};...:P_{M}\right\}  $ the
family of projects and by $\mathcal{O}$\textit{ }$=\left\{  O_{1}%
;...;O_{N}\right\}  $ the family of organizations. Projects are understood as
labeled sets of organizations and organizations as labeled sets of projects.
The corresponding intersection-graphs are denoted by $G_{P}$ and $G_{O}$ and
we will sometimes use the terms P-graph and O-graph for them. 
The size
$\left\vert x\right\vert $ of a vertex $x$ from $G_{P}$ or $G_{O}$ is the
cardinality of the set corresponding to the vertex; in the picture
of bipartite graphs, the size is just the degree of the vertex. 
In \tbls{tbl:onetworkproperties} \andtbl{tbl:pnetworkproperties}, 
we give
some basic parameters measured on the P- and O-graphs from the four
Framework Programs. Since the degree distribution
for P-graphs is a superposition of two power-law distributions (one for small
degree values and one for large values), we give the corresponding values for
the exponents parenthetically.

%: The Giant Table
\begin{table}[tbp] \centering
\begin{tabular}
[c]{|l|l|l|l|l|}\hline
\textbf{graph characteristic } & \textbf{FP1} & \textbf{FP2} & 
\textbf{FP3} & \textbf{FP4} \\\hline
\# vertices: $N$ & 2500 & 6135 & 9615 & 20873 \\\hline
( $N$ for larg.\ comp.) & (2038) & (5875) & (8920) & (20130) \\\hline
$N$ outside larg.comp. & 462 & 260 & $ $695 & 743 \\\hline
\# edges: $M$ & 9557 & 64300 & 113693 & 199965 \\\hline
(\# edges $M$ larg.comp.) & (9410) & (64162) & (113219) & (199182) \\\hline
mean degree: $\bar{d}$ & 7.65 & 20.96 & 23.65 & 19.16 \\\hline
($\bar{d}$ larg.comp.) & (9.23) & (21.84) & (25.39) & (19.79) \\\hline
maximal degree: $d_{\max}$ & 140 & 386 & 648 & 649 \\\hline
mean triangles per vertex: $\triangle$ & 22.90 
& 169.70 & 244.91 & 146.04 \\\hline
($\triangle$ larg.comp.) & (27.97) & 177.16 & 263.84 & 151.26 \\\hline
maximal triangle-number & 966 & 5295 & 15128 & 10730 \\\hline
cluster coefficient: $\bar{C}$ & 0.57 & 0.72 & 0.72 & 0.79 \\\hline
( $\bar{C}$ larg.\ comp.) & (0.67) & (0.74) & (0.75) & (0.81) \\\hline
number of components & 369 & 183 & 455 & 467 \\\hline
diameter of largest component & 9 & 7 & 9 & 10 \\\hline
mean path length: $\lambda$ of l.c. & 3.70 & 3.27 & 3.32 & 3.59 \\\hline
exponent of degree distribution & -2.1 & -2.0 & -2.0 & -2.1 \\\hline
variance of degree exponent & 0.4 & 0.3 & 0.3 & 0.3 \\\hline
exponent of org-size distr. & -2.1 & -1.9 & -1.7 & -1.8 \\\hline
variance of size exponent & 0.5 & 0.3 & 0.5 & 0.3 \\\hline
mean \# projects per org: $\mathbb{E}\left( \left\vert O\right\vert \right) $ 
& 2.40 & 4. 87 & 5.6 & 6.24 \\\hline
maximal size ($\max\left\vert O\right\vert $) 
& 130 & 82 & 138 & 172 \\\hline
\end{tabular}

\caption{Basic network properties of FP1--4 organizations projection. 
\label{tbl:onetworkproperties}}
\end{table}

\begin{table}[tbp] \centering
\begin{tabular}
[c]{|l|l|l|l|l|}\hline
\textbf{graph characteristic } & \textbf{FP1} & \textbf{FP2} & 
\textbf{FP3} & \textbf{FP4} \\\hline
\# vertices: $N$ & 3283 & 3884 & 5528 & 9087\\\hline
( $N$ for larg.\ comp.) & (2764) & (3662) & (5027) & (8566)\\\hline
$N$ outside larg.comp. & 519 & 222 & 501 & 521\\\hline
\# edges: $M$ & 51217 & 94527 & 202358 & 348542\\\hline
(\# edges $M$ larg.comp.) & (50940) & (94471) & (202306) & (348474)\\\hline
mean degree: $\bar{d}$ & 31.20 & 48.68 & 73.20 & 76.71\\\hline
($\bar{d}$ larg.\ comp.) & (36.86) & (51.60) & (80.49) & (81.36)\\\hline
maximal degree: $d_{\max}$ & 282 & 387 & 917 & 771\\\hline
mean triangles per vertex: $\triangle$ & 774.41 & 871.19 
& 1970.30 & 2034.31\\\hline
($\triangle$ larg.comp.) & 919.53 & 923.98 & 2167.05 & 2158.03\\\hline
maximal triangle-number & 12903 & 11125 & 37247 & 41141\\\hline
cluster coefficient: $\bar{C}$ & 0.67 & 0.54 & 0.44 & 0.47\\\hline
( $\bar{C}$ larg.comp.) & (0.75) & (0.57) & (0.48) & (0.50)\\\hline
number of components & 369 & 183 & 455 & 467\\\hline
diameter of largest component & 9 & 7 & 10 & 9\\\hline
mean path length: $\lambda$ of l.c. & 3.24 & 2.80 & 2.72 & 2.80\\\hline
exponent of degree distribution & (-0.8, -3.4) & (-0.7, -3.3) 
& (-0.6, -3.7) & (-0.3, -2.2)\\\hline
variance of degree exponent & (0.4, 3.6) & (0.3, 1.7) & (0.3, 1.4) & (0.2, 0.6)\\\hline
exponent of proj-size distr. & -3.59 & -2.9 & -3.2 & -4.1\\\hline
variance of size exponent & 0.6 & 0.4 & 0.2 & 0.3\\\hline
mean \# orgs per project: $\mathbb{E}\left( \left\vert P\right\vert \right) $ 
& 3.15 & 3.08 & $ $3.22 & 2.71\\\hline
maximal size ($\max\left\vert P\right\vert $) 
& 20 & 44 & 73 & 54\\\hline

\end{tabular}

\caption{Basic network properties of FP1--4 projects projection. 
\label{tbl:pnetworkproperties}}
\end{table}

As expected, FP1--4 are of small world type: high clustering coefficient and
small diameter of the giant component. There is a slight increase in the
clustering coefficient of the O-graphs from FP1 to FP4, indicating a stronger integration
amongst groups of collaborating organizations. This is also reflected in the
mean project size which increases from 2.4 to 6.2. There is an interesting
jump in the P-graph mean degree values and the mean triangle numbers between
FP1 and 2 and between FP2 and 3. The maximal degree of the O-graphs are very high in
comparison with the mean degree, which is a consequence of the power law degree
structure. For the P-graphs, the gap between mean and maximal degree is less
pronounced.

More information is
contained in the statistical properties of the relevant distributions. The
numerical data strongly indicate that the size distributions follow power
laws. Also, the O-graph degree distribution is of power-law type, while the
project-graph degree distribution is a superposition of two scale free
distributions, one dominating the distribution for small degree values (up to
100) and one relevant for the large degree values. We discuss these properties
at greater length in the following sections.

\subsection{Size distributions}

\begin{figure}
	\centering
	\includegraphics{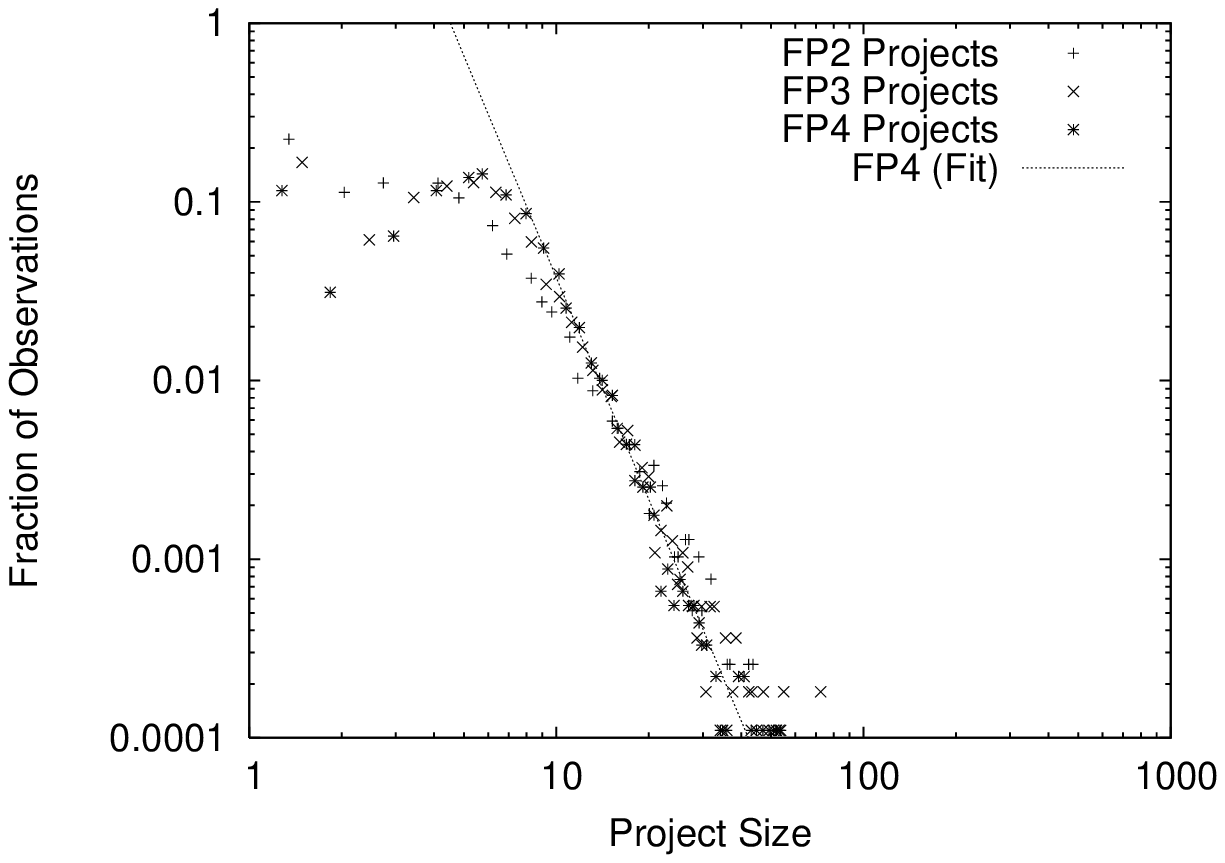}
	\caption{Distribution of project sizes.}
	\label{fig:projsize}
\end{figure}

\begin{figure}
	\centering
	\includegraphics{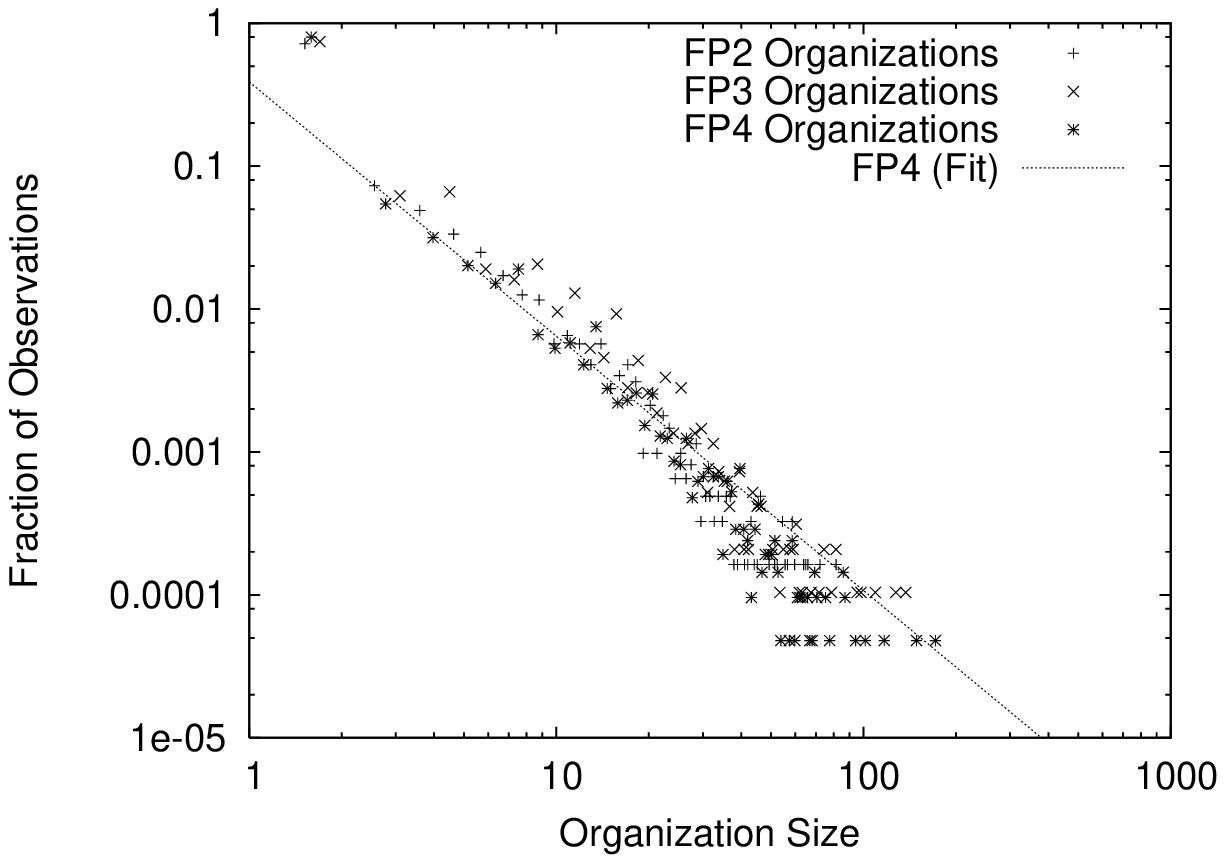}
	\caption{Distribution of organization sizes.}
	\label{fig:orgsize}
\end{figure}

The size distributions are the basic distributions for the EU-networks since,
as will be shown in section \sxn{sec:molloyreed}, a typical sample from the
random graph space with fixed size distributions like in FP 2-4 will have very
similar statistical properties to FP 2-4. This strongly suggests that there is
essentially no additional correlation in the data once the size distribution
is known. 
Both the O-graph and P-graph size distributions show clear asymptotic power
law distributions for FP1--4 (\figs{fig:projsize} \andfig{fig:orgsize}). In terms of the corresponding bipartite
graph, these are just the degree distributions of the project and organization partitions. 
While the O-graph size distribution is of power law type over the
whole size range, the P-graph size distribution deviates
strongly from the power-law for small size values. In \sxn{sec:randgraphmodel},
we give a possible
explanation for the appearance of the power law distribution for size. 

The numerical values for the exponents of the organization size distributions
from FP2--4 are slightly below 2, but constant within the error tolerance. This
indicates that the distribution of organizations able to carry out a
particular number of projects has not changed in the three Framework Programs.
A complementary interpretation of this finding is that the underlying research
activities, which we know to have changed over time, have not altered the mix
of organizations participating in a particular number of projects in each
Framework Program. It is further worth noting that the values of the O-graph
exponents are close to the  critical value 2, hence the size expectation could
diverge for large graphs (whether the value is really below 2 or not is 
still unclear due to the error tolerance ). 

The picture is similar for the P-graphs,
although there are some differences in the initial behavior (that is, for
small project sizes) and in the exponent value. The local minima at size 2 is
decreasing from FP2--4. This points to the existence of an optimal project
size within the regime of the EU FPs. Moreover, the rise in the average
project size indicates that increases in the available funding from FP2 to FP4
not only lead to more projects, but also slightly larger projects. This is
consistent with recommendations from evaluation studies and the stated
attempts of the EU commission to reduce its administrative burden. As a whole,
the size distribution for the P-graphs matches in the asymptotic  regime very
well to a power law with exponent around -3, hence indicating that the
mechanisms for coagulation of organizations into a project did not greatly
change from FP2--4.

\subsection{The degree distribution}

Since the degree distribution in the projection graphs is just the
distribution of the size of the 2-neighborhood $N_{2}\left(  x\right)
:=\#\left\{  y:d^{bi}\left(  x,y\right)  =2\right\}  $, it is not surprising
that this quantity is closely connected to the size distribution. In the
absence of other special correlations, it can be shown (see
\sxn{sec:randgraphmodel}) that the degree distribution is determined by the
size distribution in a rather simple way. Namely, for the case when both size
distributions are scale-free with exponents, say $\alpha$ (O-size) and $\beta$
(P-size), the P-graph degree distribution is a superposition of two power-law
distributions with exponents $\alpha-1$ (and cutoff given by the maximal
O-size value) and $\beta$. The same holds vice versa for the O-graph.

\begin{figure}
	\centering
	\includegraphics{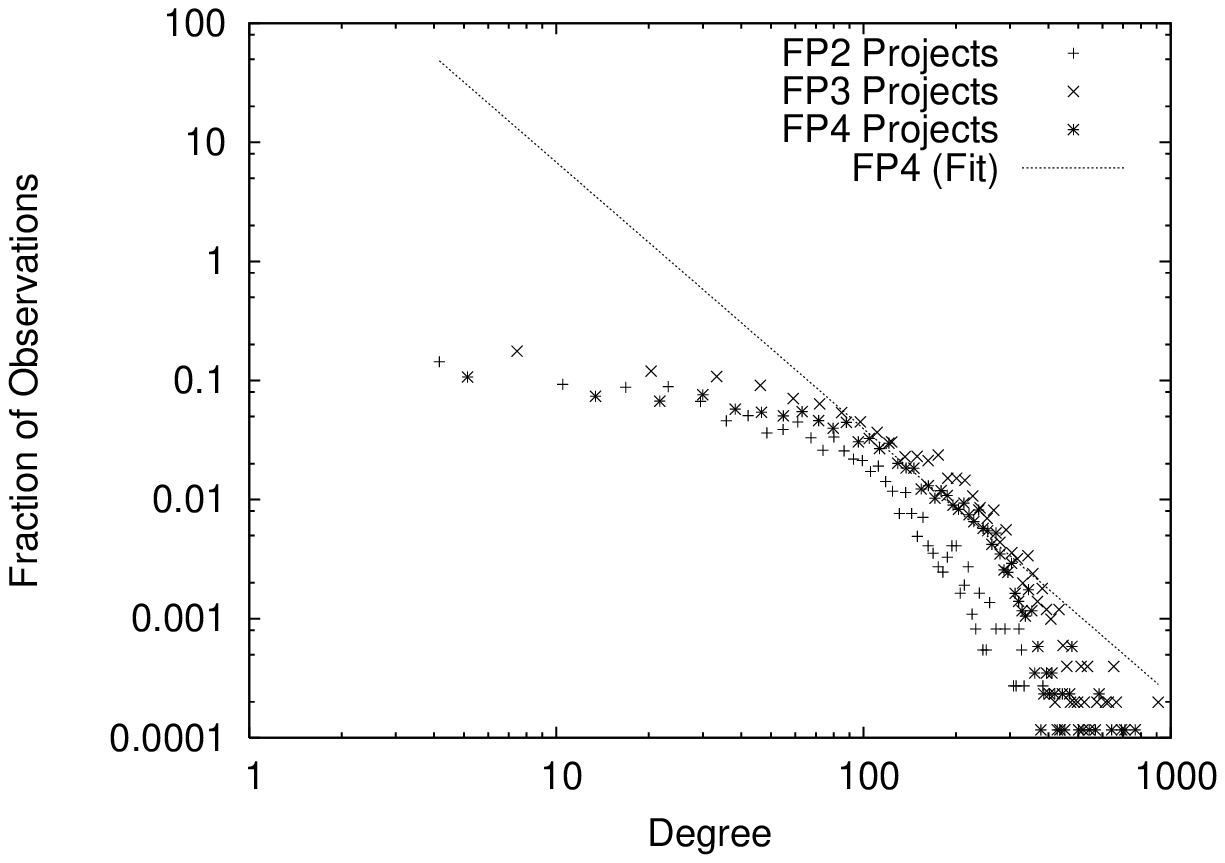}
	\caption{Degree distribution of projects projection.}
	\label{fig:projdegrees}
\end{figure}

\begin{figure}
	\centering
	\includegraphics{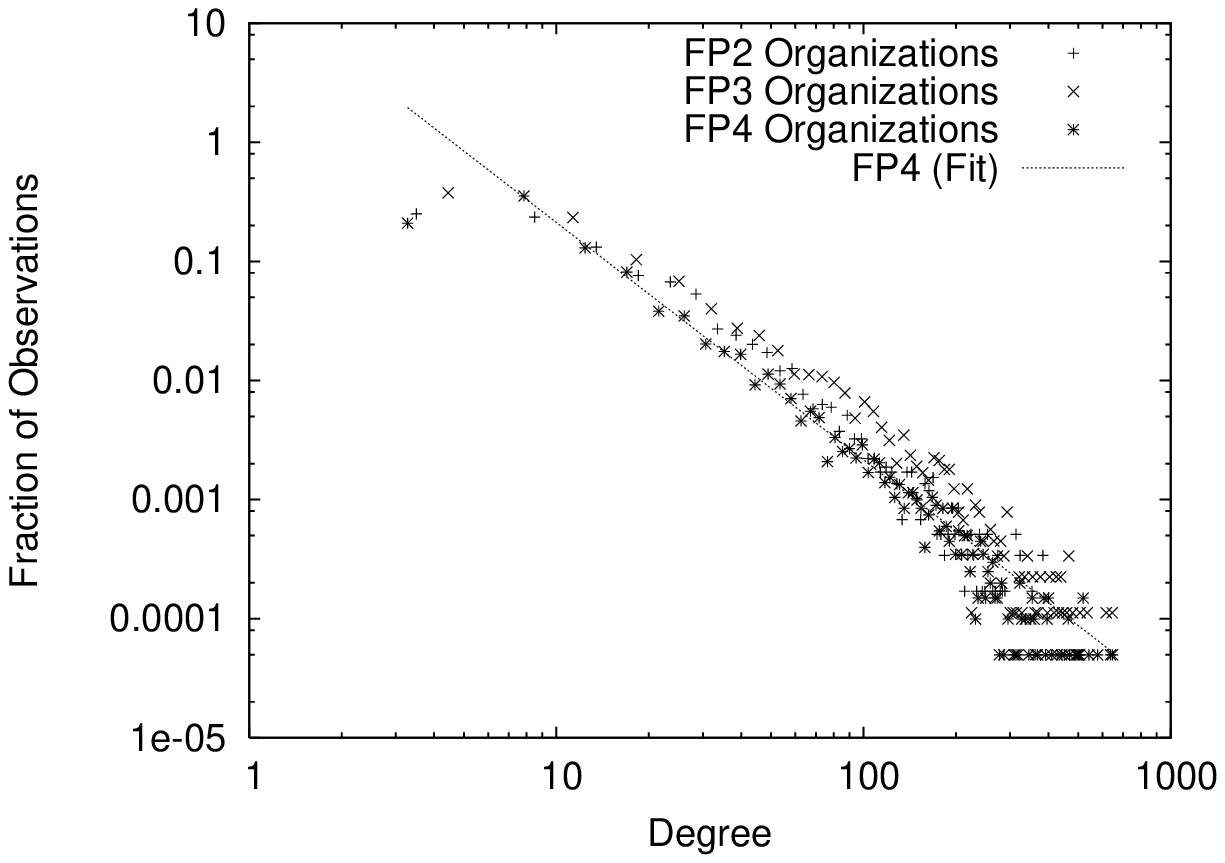}
	\caption{Degree distribution of organizations projection.}
	\label{fig:orgdegrees}
\end{figure}

In \figs{fig:projdegrees} \andfig{fig:orgdegrees}, we show the
degree-distribution for the P- and O-graphs in a log-log plot. . While the
organization graphs for FP2--4 show a clear power law, the picture for the
project graphs is more complicated. As previously mentioned, the P-graph degree
distribution shows two different power laws, one for the initial segment up to
degree 150 and another one for large degrees. Nevertheless, there is still a
widely scattered heavy tail in the degree distribution. The deviation from a
power law in the P-graphs indicates a kind of anticorrelation: large projects
above a size of 15 are mainly formed by organizations of small size. A
possible explanation is that large projects have a time- and
resource-demanding intrinsic network structure, making it more unlikely that a
participating organization has other projects (of course, with the exception
of hub-like organizations such as CNRS with \apriori\ unlimited capacity).

\subsection{Clustering, correlation and edge multiplicity}

\begin{figure}
	\centering
	\includegraphics{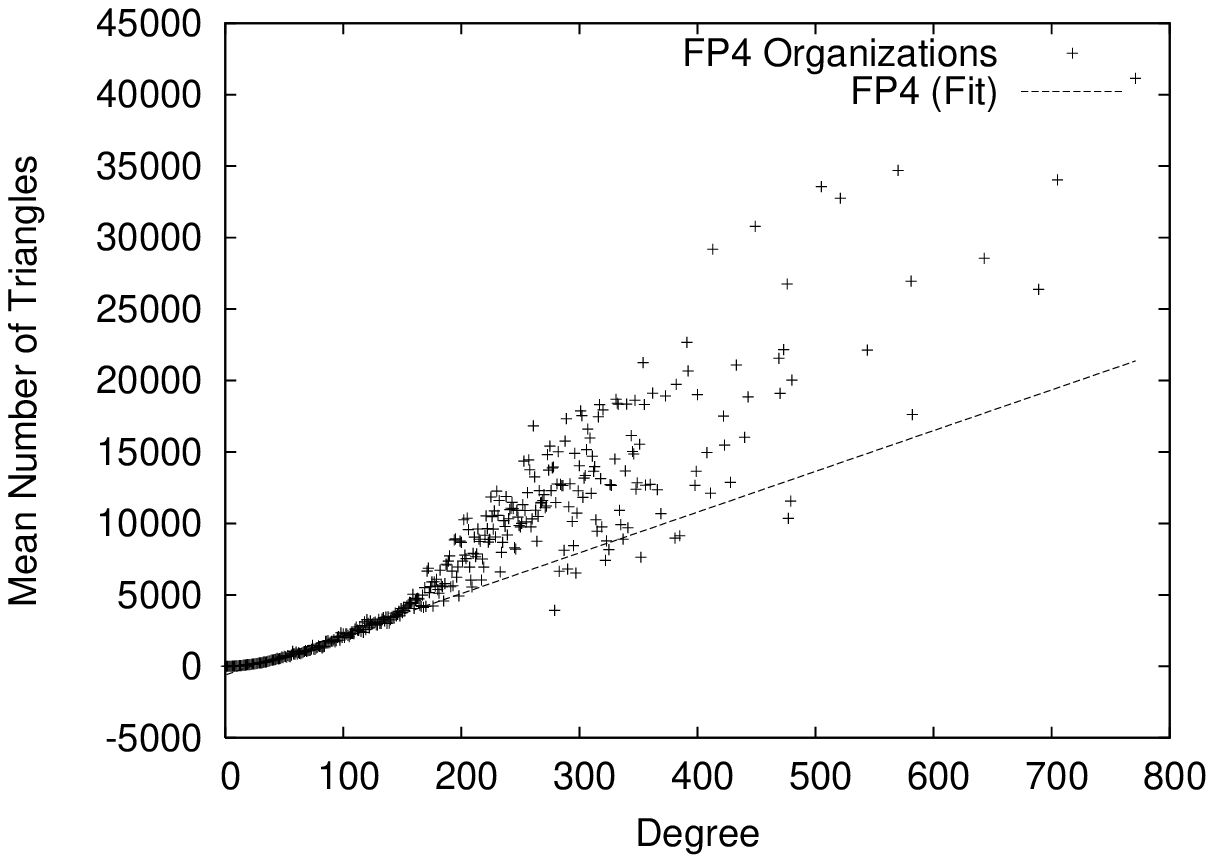}
	\caption{Relation between degree and number of triangles in the projects projection.}
	\label{fig:projdegreetris}
\end{figure}

\begin{figure}
	\centering
	\includegraphics{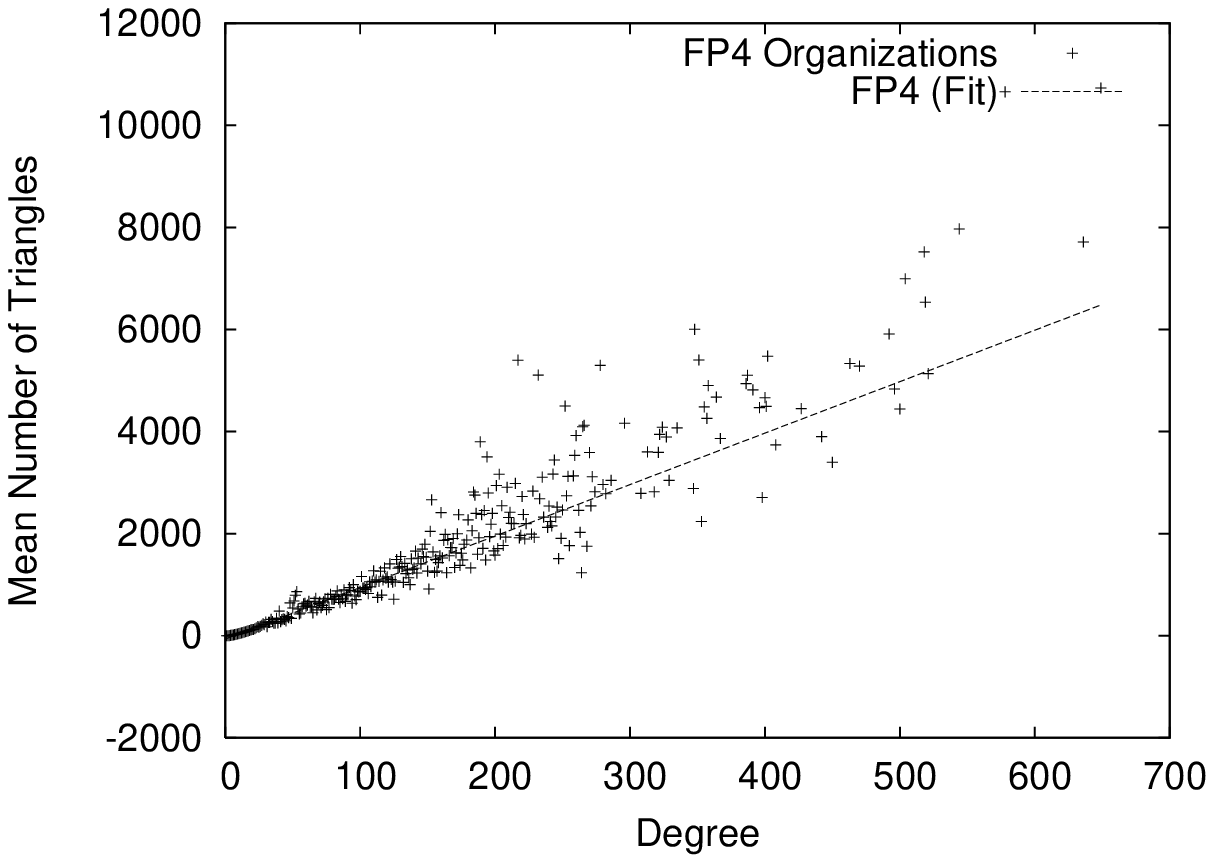}
	\caption{Relation between degree and number of triangles in the 
	organizations projection.}
	\label{fig:orgdegreetris}
\end{figure}

By their construction process, intersection graphs have a naturally high
clustering coefficient. This is easily seen, since an organization which
participates in, say, $k$ projects generates a complete subgraph of order $k$
in the P-graph amongst these projects. If the probability for an organization
to be in more than one project is asymptotically bound away from zero, it
follows that the P-graph (and similarly for the O-graph through an analogous
argument) has a nonvanishing clustering coefficient. In the present study, we
focus on the triangle number $\triangle\left(  x\right)  :=\#\left\{
\text{triangles containing }x;x\in\left(  \mathcal{P}\text{ or }%
\mathcal{O}\right)  \right\}  $ as a measure of local clustering. We define
the degree-conditional mean triangle number as $\triangle_{k}:=\mathbb{E}%
\left\{  \triangle\left(  x\right)  \mid d\left(  x\right)  =k\right\}  $. As
seen in \figs{fig:projdegreetris} \andfig{fig:orgdegreetris}, we have
$\triangle_{k}\sim k$ for both graph types.

There is a good explanation for this type of behavior in the framework of
intersection graphs (see \sxn{sec:randgraphmodel}). As noted above, high
clustering in intersection graphs is not necessarily an indication of local
correlations between vertices. This is already seen in the case of an
Erd\"{o}s-Renyi random bipartite graph where an edge between any project and
organization is drawn i.i.d.\ with probability $p$. If $\mathcal{P}$ and
$\mathcal{O}$ are of equal cardinality $N$ and $p=\frac{c}{N}$, the expected
bipartite degree equals $c$. For large $N$ a typical realization of the random
graph looks locally like a tree with branching number $c-1$. However, for the
projection graphs, we obtain an positive clustering coefficient that is
independent of $N$, since most projects and organizations cause complete
graphs of order $c$ and a typical vertex is therefore a member of $\sim c$
cliques of order $c$. 

\begin{figure}
	\centering
	\includegraphics{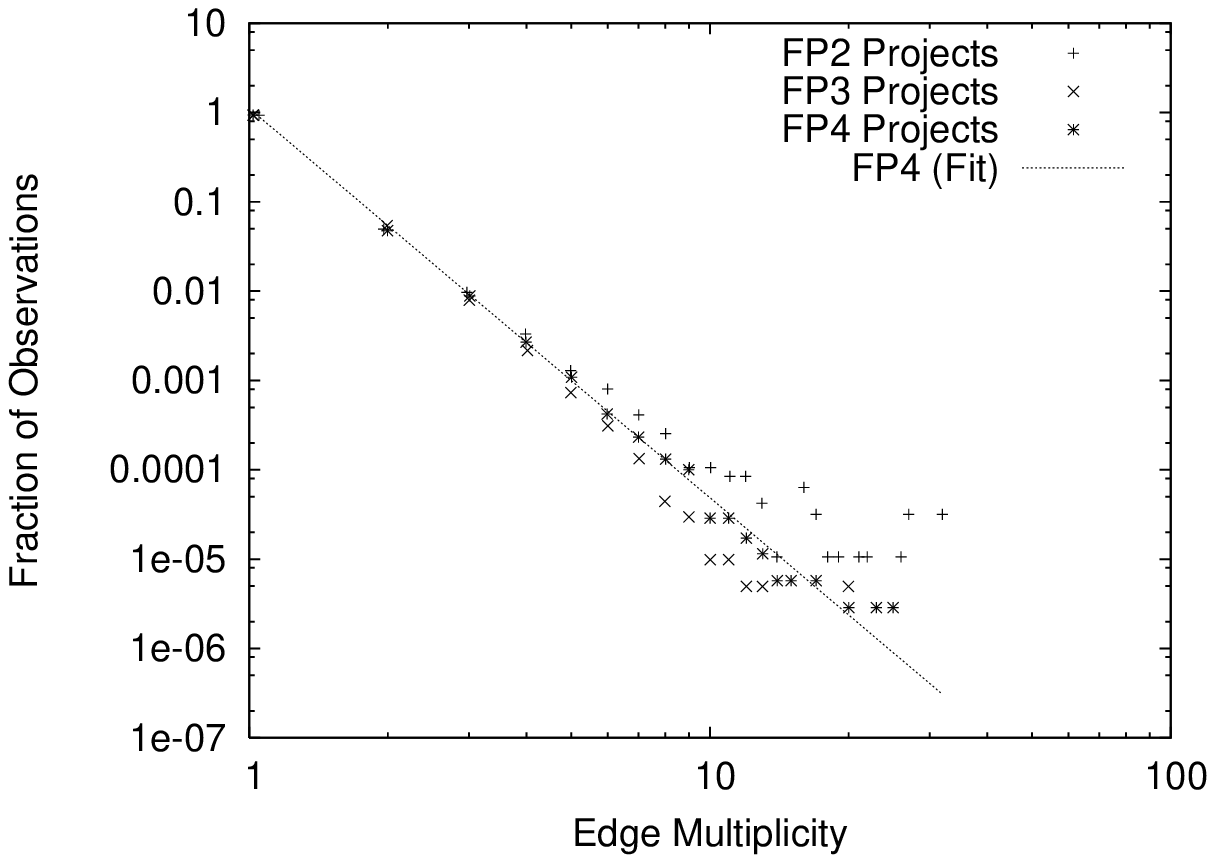}
	\caption{Distribution of edge multiplicities in the projects projection.}
	\label{fig:projedgemult}
\end{figure}

\begin{figure}
	\centering
	\includegraphics{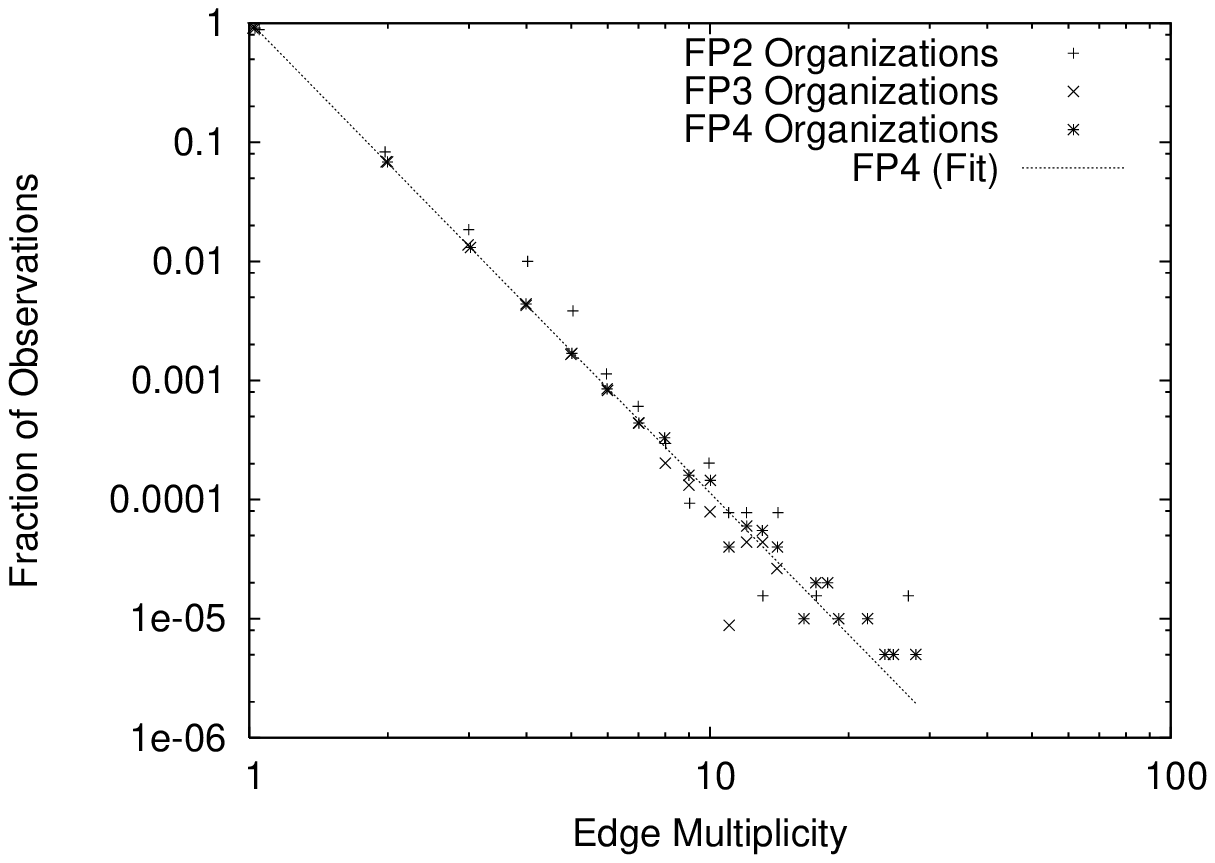}
	\caption{Distribution of edge multiplicities in the projects projection.}
	\label{fig:orgedgemult}
\end{figure}

A better indication for the presence of correlations is
given by the so-called multiplicity of edges. For a link between two
organizations or projects it is sufficient to have just one project or
organization, respectively, in common, but of course there could be more.
Given an edge $x\sim y$, we define $m\left(  x,y\right)  :=\left\vert x\cap
y\right\vert -1$ and call it the multiplicity of the edge. As will be
discussed in the next section, random intersection graphs without local search
rules can nevertheless admit a high edge multiplicity. In
\fig{fig:projedgemult} \andfig{fig:orgedgemult}, the multiplicity distribution
is shown for P- and O-graphs of FP2--4. There is an almost perfect power-law
behavior with exponent 4.3. Note that positive multiplicity in the projection
graphs translates in the bipartite graph picture into the presence of cycles
of length four. The presence of exceptionally high multiplicity in the
P-graphs may be caused by memory effects due to prior collaborative
experience. Also, a greater edge multiplicity may result from the fact that
organizations are active in a wider set of complementary activities. In this
case, intra-organizational spillovers may also be of importance as search for
potential partners may be influenced by the collaboration behavior of other
actors within an organization. Such effects should be detectable from a fine
structure analysis of the time evolution of the corresponding graphs. 

\subsection{Diameter and mean path length}

There is essentially no difference in the diameter value of the largest
component in the four Framework Program networks. 
A classical random graph of the same 
size and the same
edge number would have a diameter about $\log_{\bar{d}}N $. % $\sim $ ??. 
The mean path
length is about a third of the diameter and and shows a slightly higher
variation between the different framework programs. It is well known that the
expected path length in random graphs with a scale free degree distribution
and exponent less than 3 is essentially independent of the graph size (the
diameter of the largest component still increases in $N$ but only as $\log\log
N$). The same holds for random intersection graphs with power law size and 
degree
distributions. Since the the O-graphs seem to fall into that class, the almost
constant diameter and path length is not surprising. Although the P-graphs
do not show an asymptotic power law structure for the degree, there is a strong
increase in the edge density from FP2 to FP4, keeping the diameter of the
largest component almost fixed.

\section{A random intersection graph model} \label{sec:randgraphmodel}

Intersection graphs are a natural framework for networks derived from a
membership relation, such as citation networks, actors networks, or networks reflecting
any other kind
of cooperation. As previously mentioned, intersection graphs by construction have
a high clustering coefficient. As explained below, the
clique distribution of a random intersection graph is almost given by the size
distribution of the dual graph.

\subsection{Random intersection graphs with given size distribution} 
\label{sec:randgraphmodelfixedsize}

One of the simplest random intersection models is constructed in the following
way. Knowing the size of a set to be constructed, we generate a random subset
from a finite base set $X=\left\{  a_{1},a_{2},...,a_{N}\right\}  $ of $N$
elements, such that each set element is drawn i.i.d. uniformly from $X$. These
subsets constitute the vertices of a random graph. Edges are defined via
the set intersection property, namely we have an edge between $i$ and $j$
(denoted by $i\sim j$) if and only if the associated subsets $A_{i}$ and $A_{j}$ have
nonempty intersection (to compare with earlier sections, $A$ stands here
for either projects sets $P$ or organization sets $O$). The size (cardinality)
of the subsets is either itself a random variable drawn i.i.d. from a
probability distribution $\varphi(k)$ or given by a list
$\left\{  D_{k}:=\#\left\{  A_{i}:\left\vert A_{i}\right\vert \right\}
=k\right\}  $ (where for each $i$ a conditional random choice is made to which
size class it belongs). For the latter case, we define again $\varphi
(k):=\frac{D_{k}}{M}$ where $M$ is the total number of sets to be formed.

Since we want to compare the model with the EU- cooperation network we are
mainly interested in the situation when $\varphi$ is an asymptotic power law
distribution
\begin{equation}
\varphi(k)=\frac{1}{k^{\alpha+o\left(  1\right)  }};\alpha>2
\mathperiod
\end{equation}
This assumption is also reasonable for many other applications where
vertices are formed from a base set of elements. To obtain an interesting
limiting random graph space, we further assume that the number of chosen
subsets is $C_{1}\cdot N$ where $C_{1}$ is neither too large nor too small (for
FP2--4 we have about twice as many organization as projects hence hence $C_{1}$
is either $2$ or $0.5$). 

A basic quantity for the analysis of intersection
graphs is the conditional edge probability given the size of two subsets:
\begin{align}
&  P_{k,l}\left(  N\right)  := 
	\Pr\left\{  i\sim j\mid\left\vert A_{i}\right\vert 
	=k\text{ and }\left\vert A_{j}\right\vert =l\text{ }\right\} \\
&  =\Pr\left\{  A_{i}\cap A_{j}\neq\emptyset\mid\left\vert A_{i}\right\vert
=k\text{ and }\left\vert A_{j}\right\vert =l\right\}  \\
& =1-\frac{\binom{N-k}{l}}{\binom{N}{l}}\\
&  =1-\frac{\left(  N-k\right)  !\left(  N-l\right)  !}{N!\left(
	N-k-l\right)  !}\\
& =1-\frac{\left(  N-k\right)  \left(  N-k-1\right)
\cdot...\cdot\left(  N-k-l+1\right)  }{N\left(  N-1\right)  \left(
N-2\right)  \cdot...\cdot\left(  N-l+1\right)  }
\mathperiod
\end{align}
Using the condition $lk\ll N$, we obtain
\begin{align}
P_{k,l}\left(  N\right)   &  =1-\frac{\left(  1-\frac{k}{N}\right)  \left(
1-\frac{k+1}{N}\right)  ...\left(  1-\frac{k+l-1}{N}\right)  }{\left(
1-\frac{1}{N}\right)  \left(  1-\frac{2}{N}\right)  ...\left(  1-\frac{l-1}%
{N}\right)  }\\
&  =1-\frac{1-\frac{lk+\frac{1}{2}\left(  l-1\right)  \left(  l-2\right)  }%
{N}+o\left(  \frac{1}{N}\right)  }{1-\frac{\frac{1}{2}\left(  l-1\right)
\left(  l-2\right)  }{N}+o\left(  \frac{1}{N}\right)  }\\
&  =\frac{lk}{N}+o\left(  \frac{1}{N}\right)
\mathperiod
\end{align}
With this result, we can easily calculate the conditional degree distribution
for a vertex of given size. First, we estimate the conditional subdegree
distribution with respect to a given group of vertices of size $m.$ Here, the
subdegree $d_{m}\left(  i\right)  $ of a vertex $i$ is defined as the number
of edges $i$ has with vertices of size $m.$ Clearly $d\left(  i\right)
=\sum\limits_{m}d_{m}\left(  i\right)  .$ We have
\begin{align}
\psi_{l}\left(  k,m\right)   &  :=\Pr\left\{  d_{m}\left(  i\right)
	=k\mid\left\vert A_{i}\right\vert =l\text{ }\right\} \\
	&  =\sum_{G}\Pr\left\{  \sharp\left\{  j\mid\left\vert A_{j}\right\vert
=m\right\}  =G\right\}  \binom{G}{k}\left(  \frac{ml}{N}+o\left(  \frac{1}%
{N}\right)  \right)  ^{k}\left(  1-\frac{ml}{N}+o\left(  \frac{1}{N}\right)
\right)  ^{G-k}
\mathperiod
\end{align}
The probability that a randomly chosen vertex $j$ has size $m$ equals, by
assumption, $\frac{C_{2}}{m^{\alpha+o\left(  1\right)  }}$ with normalization
constant $C_{2}$ ( $1=\sum\limits_{m}\frac{C_{2}}{m^{\alpha+o\left(  1\right)
}}$ ). We therefore obtain
\begin{equation}
\psi_{l}\left(  k,m\right)  =\lim\limits_{N\rightarrow\infty}\binom
{C_{1}N\cdot\frac{C_{2}}{m^{\alpha}}}{k}\left(  \frac{ml}{N}+o\left(  \frac
{1}{N}\right)  \right)  ^{k}\left(  1-\frac{ml}{N}+o\left(  \frac{1}%
{N}\right)  \right)  ^{C_{1}N\cdot\frac{C_{2}}{m^{\alpha}}-k}
\mathcomma
\end{equation}
which converges to a Poisson distribution
\begin{equation}
\psi_{l}\left(  k,m\right)  =\frac{c\left(  m\right)  ^{k}}{k!}e^{-c\left(
m\right)  }%
\end{equation}
with $c\left(  m\right)  = m^{1-\alpha}lC_{1}C_{2}.$ Since the distribution
$\psi_{l}\left(  k\right)  $ of the degree of vertices $i$ with $\left\vert
A_{i}\right\vert =l$ is the convolution of the 
Poisson distributions $\psi_{l}\left(  k,m\right)  $, we obtain again a Poisson 
distribution for $\psi_{l}\left(  k\right)  $ :
\begin{equation}
\psi_{l}\left(  k\right)  =\frac{c_{l}^{k}}{k!}e^{-c_{l}}%
\end{equation}
with $c_{l}=\sum\limits_{m}c\left(  m\right)  =l\cdot C_{3}$, where $C_{3}%
=\sum\limits_{m}m^{1-\alpha}C_{1}C_{2}$ is a well defined constant since
$\alpha>2.$ It remains to estimate the total degree distribution $\psi\left(
k\right) $. In \cite{6}, conditions were given describing when a superposition
of Poisson distributions results in a scale-free distribution. Specifically, we
get the following asymptotic estimate:%
\begin{align}
\psi\left(  k\right)   &  =\sum\limits_{m}\varphi\left(  m\right)
\frac{\left(  mC_{3}\right)  ^{k}}{k!}e^{-mC_{3}}\\
&  =\sum\limits_{m}\frac{1}{m^{\alpha+o\left(  1\right)  }}\cdot\frac{\left(
mC_{3}\right)  ^{k}}{k!}e^{-mC_{3}}%
\mathperiod
\end{align}
The main contribution to $\psi\left(  k\right)  $ comes from a rather small
interval of $m$-values,  called $I_{ess}\left(  k\right)  $. This interval has  the
property that for $m\in I_{ess}\left(  k\right)  $, the expectation
$\mathbb{E}\left(  d\left(  i\right)  \mid\left\vert A_{i}\right\vert
=m\right)  $ is of order $k.$ The exponential decay of the Poisson
distribution guarantees that the remaining parts of the sum become
arbitrarily small for large $k$. It is important that the constant $c_{l}$ has a
linear $l-$dependence since an $l-$proportionality with exponent larger than one
would force the degree distribution to have gaps due to a lack of overlap of
the individual Poisson distributions. We therefore obtain for the degree
distribution a power law with the same exponent $\alpha$ as in the size distribution.

Although the intersection model gives a power-law degree distribution 
when the size distribution is already of power-law type, we will not obtain a
power-law distribution for the size on the dual graph unless additional
assumptions are made on the set formation rules. It is easy to see that the
size distribution on the dual graph is asymptotically Poisson. Since
$\Pr\left\{  \left\vert x\right\vert =k\right\}  \sim$ $\binom{M}{k}\left(
\frac{\mathbb{E}\left(  \left\vert A\right\vert \right)  }{N}\right)
^{k}\left(  1-\frac{\mathbb{E}\left(  \left\vert A\right\vert \right)  }%
{N}\right)  ^{M-k}$ and $\mathbb{E}\left(  \left\vert A\right\vert \right)  $
converges as well as $\frac{M}{N}$ for $M,N\rightarrow\infty$, we obtain in the
limit a Poisson distribution. Nevertheless, the degree distribution on the dual
graph still admits a scale-free part induced by the scale-free size
distribution of the intersection graph. We will not discuss many of the details, but
instead
provide a simple estimation for the lower bound on the number of elements
$a_{i}$ with $d\left(  a_{i}\right)  =k$. Namely, the number of elements
$a_{i}$ which are members of sets $A_{j}$ with $\left\vert A_{j}\right\vert
=k$ is for large $k$ and $M,N>>k$ about $\frac{k\cdot M\cdot const}{k^{\alpha
}}=\frac{N\cdot const}{k^{\alpha-1}}$. Since $d\left(  a_{i}\right)  \geq k$
for $a_{i}\in A_{j}$ with $\left\vert A_{j}\right\vert =k$, we obtain
$\frac{const}{k^{\alpha-2}}$ as a lower bound on the density of elements
$a_{i}$ with degree greater than or equal to $k$ (note that we assumed $\alpha>2$). This
estimate holds of course only up to the maximal size value $k$, which is in
the range of the power law distribution for the set sizes $\left\vert
A_{i}\right\vert .$ For larger $k$-values there is a rapid exponential decay.

The last argument clarifies also the situation when one wants to impose
conditions on the size distribution and the dual size distribution. Without
going into the details of the rather involved analysis, we simply state that the
resulting degree distribution is given by a superposition of the size
distibution and the dual size distribution (the last one enters with an
exponent reduced by one). This explains essentially the picture for the degree
distribution for the P-graph. 

Finally we want to discuss the mean triangle (conditioned on the degree) -
degree dependence which shows a clear linear behavior in the empirical data.
We argue that this is again a consequence of the power law distribution
for the size. First observe that a size $k$ element $a_{i}\in A_{j}$ induces a
$k-1$ complete subgraph on the neighborhood vertices of $A_{j}$. Furthermore,
each maximal $k-$clique in which $A_{j}$ is a member generates $\left(
k-1\right)  \left(  k-2\right)  /2$ triangles for $A_{j}$. Since the size
distribution of the elements $a_{i}$ is Poisson with expectation of, say, $c$ and
the degree of $A_{j}$ is proportional to the 
size $\left\vert A_{j}\right\vert$, 
we obtain for the conditional expected number of triangles $\triangle_{k}$
given the degree $k$:%
\begin{equation}
\triangle_{k}:=\mathbb{E}\left(  \#\text{triangles containing }A\mid d\left(
A\right)  =k\right)  \sim\frac{c^{2}}{2}const\cdot k
\label{eq:condexpectedtrisgivendegree}
\mathperiod
\end{equation}
In deriving \eqn{eq:condexpectedtrisgivendegree}, we used the facts that with high probability
the size of the intersection between two sets $A_{i}$ and $A_{j}$ has
cardinality $1$ (conditioned on the two sets having a nonempty intersection)
and that the Poisson distribution has an exponentially decaying tail.

\subsection{A Molloy-Reed version of random intersection graphs and a
Bernoulli type model}\label{sec:molloyreed}

We sketch the construction of random intersection graphs with given size
distribution $\varphi$ and size distribution $\psi$ on the dual. The two
distributions are not independent but have to fulfill the condition 
$\sum_{i}\left[  \varphi\left(  i\right)  -\psi\left(  i\right)  \right]  =0$. There
are further restrictions on the maximal size in order to get a reasonable random graph
model. Note that the problem is equivalent to the construction of a random
bipartite graph given the degree sequence on the two partitions. 

Assign first
to each set $A$ and each element $a$ from the base set a random size value
according to the given distributions $\varphi$ and $\psi$. Let $D_{k}$ be the
resulting set of elements $a_{i}$ with size $k$. Replace each element from
$D_{k}$ by $k$ virtual elements $a_{i,l}, l=1, 2, \ldots, k$ and form a new base set
$X^{\prime}$ with all the virtual elements. The set formation process for the
sets $\left\{  A_{i}\right\}  $ is now the same as in the previous section
except that each chosen virtual element $a_{i,l}$ will be removed from
$X^{\prime}$ when it was selected first into a set. After the sets are
constructed we identify the virtual elements back into the original ones and
define the corresponding set graph in the usual way.

By construction the
resulting size distribution on the dual graph will be given by $\psi$ as long
as the probability of choosing two virtual elements $a_{i,l}$ and $a_{i,m}$
(corresponding to the same element $a_{i}$) is sufficiently small. To ensure
this one has to impose restrictions on the maximal size values. It is not
difficult to show that the correlation between the size of $A$ and the size of
an element $a$ is multiplicative. In case of a linear relation between the
number of sets $N$ and the number of elements $M$ we have
\begin{equation}
\Pr\left\{  a\in A\mid\left\vert A\right\vert =k\wedge\left\vert a\right\vert
=l\right\}  \sim\frac{const}{N}k\cdot l
\mathperiod
\end{equation}
To see this\ observe that%
\begin{align}
\Pr\left\{  a\in A\mid\left\vert A\right\vert =k\wedge\left\vert a\right\vert
=l\right\}   &  =1-\Pr\left\{
\begin{array}
[c]{c}%
\text{among the }k\text{ choices to generate }A\\
\text{ is no virtual }a-\text{ element}%
\end{array}
\right\}  \\
&  =1-\frac{M^{\ast}-l}{M^{\ast}}\cdot\frac{M^{\ast}-1-l}{M^{\ast}-1}%
\cdot...\cdot\frac{M^{\ast}-k-l+1}{M^{\ast}-k+1}%
\end{align}
with $M^{\ast}$ being the number of virtual elements. The last formula has the
same structure as the expression for the pairing probability in the previous
section hence we get, for $lk\ll M^{\ast}$ and bounded first moments of the
$\psi$-distribution, the claimed multiplicative correlation. We note that
there is also a variant of the Molloy-Reed construction which produces an
additive size-size correlation such that $\Pr\left\{  a\in A\mid\left\vert
A\right\vert =k\wedge\left\vert a\right\vert =l\right\}  \sim\frac{const}%
{N}\left(  k+l\right)  $ holds (see \cite{9} for details of the algorithm).

\begin{figure}
	\centering
	\includegraphics{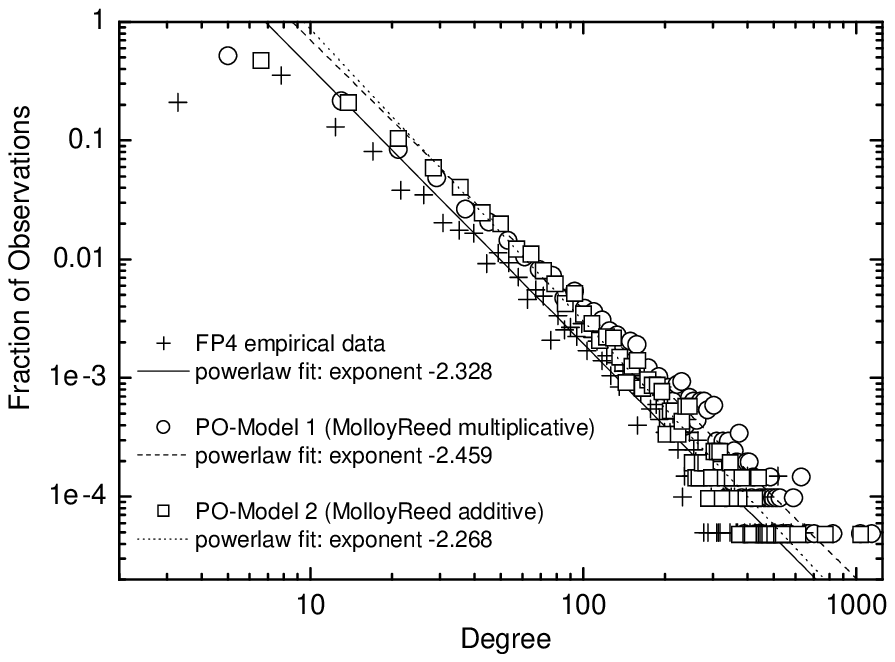}
	\caption{Degree distribution for the O-graphs.}
	\label{fig:odegree}
\end{figure}

\begin{figure}
	\centering
	\includegraphics{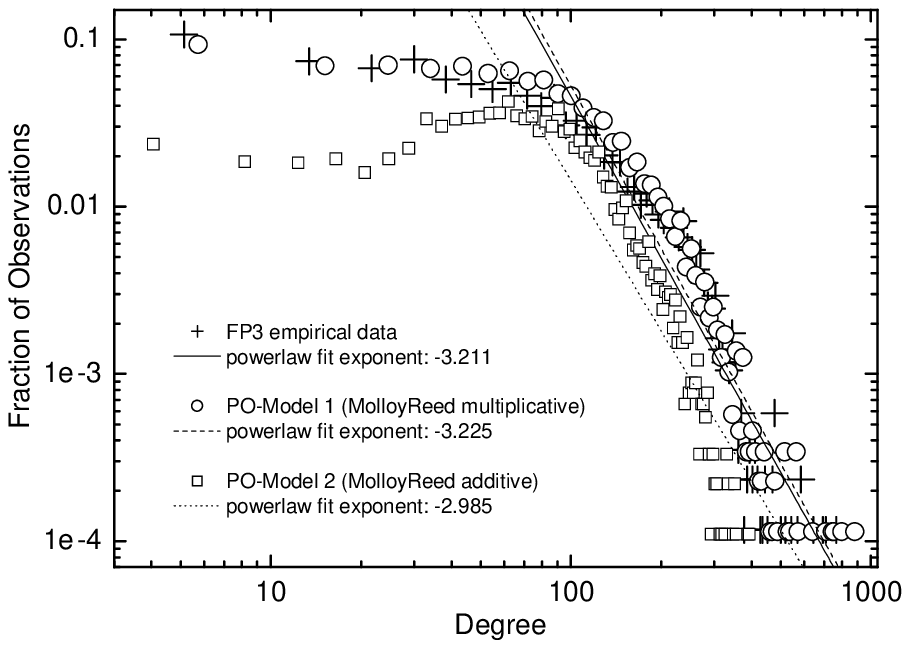}
	\caption{Degree distribution for the P-graphs.}
	\label{fig:pdegree}
\end{figure}

We next present a simulation-based comparison of the multiplicative and
additive Molley-Reed model with the FP4 network. The input size distributions
for the Molloy-Reed simulations are the same as in FP4. For completeness we
also include the simulation results based on the simple random intersection
graph model defined in the previous section. To make clear which size
distribution is given in that case we use the notation P-model (O-model) for
the intersection graph with fixed P (O) size distribution and denote by
PO-model the corresponding Molloy-Reed graphs since both size distributions
are fixed therein. \Figs{fig:odegree}
\andfig{fig:pdegree} show the degree distribution for
the O- and P-graphs. There is a very good agreement over the whole range of
degree values between the real FP4 network projections and typical samples of
the multiplicative Molloy-Reed model.  This is quite remarkable since a
considerable bias from the almost independence of the Molloy-Reed model should
be visible in the degree distributions. The fact that there is no deviation
between the degree distributions indicates that the majority of 
project-organization alignments is essentially a random process. Furthermore, the
additive model reproduces the FP4 P-graph degree distribution only well for
large degree values indicating that the correlation is indeed multiplicative.

\begin{figure}
	\centering
	\includegraphics{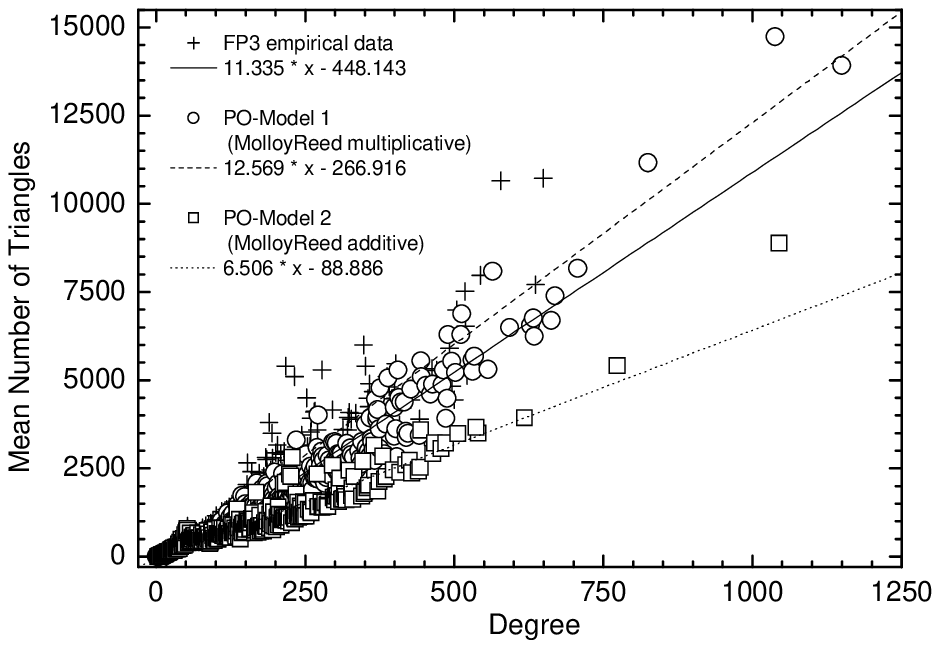}
	\caption{Triangle-degree correlation for the O-graphs.}
	\label{fig:odegtris}
\end{figure}

\begin{figure}
	\centering
	\includegraphics{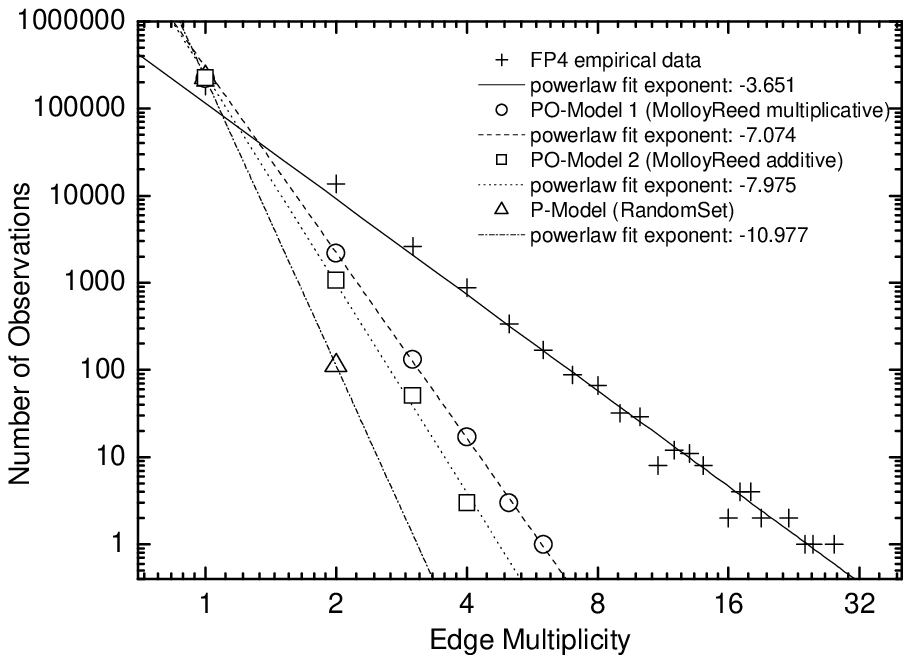}
	\caption{Edge multiplicity for the O-graphs.}
	\label{fig:oedgemult}
\end{figure}

Two quantities measuring local correlations are the triangle-degree dependence
and the distribution of edge multiplicity introduced earlier. \Fig{fig:odegtris}
compares the triangle-degree correlation for the O-graph. Although the overall
picture is similar (linear dependence up to medium degree) there is a clear
tendency for higher triangle numbers in FP4 for large degree values. Again the
multiplicative version matches better with the data then the additive model.
The edge multiplicity---again for the O-graphs---is shown in \fig{fig:oedgemult}. The
real graph has a considerably smaller value in the exponent and extends to
almost twice as large a maximal multiplicity value. Nevertheless, both
Molloy-Reed models show a sharp scale-free distribution for the multiplicity.
This is quite surprising, since, naively, one would expect the probability for
positive edge multiplicity to go to zero as $N$ becomes large. In summary, one
has a strong agreement between the real data and the multiplicative
Molloy-Reed model (the comparison results for FP2 and FP3 are almost identical
to the situation with FP4 and have therefore not been depicted here). Only in
the fine structure of clustering characteristics are some differences observed.

Finally, we briefly outline why, under certain
circumstances, almost independent models like the Molloy-Reed one can have a
scale-free edge-multiplicity distribution. To keep the discussion as
transparent as possible, we study the question in a pure bipartite Bernoulli
model, which can be thought of as a kind of predecessor to the Cameo-model
discussed below.

To each vertex from the O- and P- partitions (with cardinality $N$ and $M$), we
assign a power-law distributed, positive integer parameter $\mu\left(
P\right)  $ and $\nu\left(  O\right)  $ with exponents $\alpha$ and $\beta$.
That is we partition the P- and O-vertices into sets $D_{\mu}:=\#\left\{
P\mid\mu\left(  P\right)  =\mu\right\}  $ \ and $G_{\nu}:=\#\left\{  O\mid
\nu\left(  O\right)  =\nu\right\}  $ such that $\left\vert D_{\mu}\right\vert
=\frac{C_{P}M}{\mu^{\alpha}}$ and $\left\vert G_{\nu}\right\vert =\frac
{C_{O}N}{\nu^{\beta}}$ where $C_{P}$ and $C_{O}$ are normalization constants.
We further assume $N=C_{op}\cdot M$ and put
\begin{equation}
\Pr\left\{  P\sim O\right\}  :=\frac{c}{N}\mu\left(  P\right)  \nu\left(
O\right)
\mathperiod
\end{equation}
It is easy to see that the expected degree, conditioned on the $\mu$ or
$\nu$ value, is proportional to $\mu$ or $\nu,$ respectively, and therefore the
(bipartite) degree distribution on each partition has the same exponent as
$\mu$ or $\nu$. Note that the maximal $\mu$ and $\nu$ values are given by
$\mu_{\max}\sim M^{\frac{1}{\alpha}}$ and $\nu_{\max}\sim N^{\frac{1}{\beta}}$. 

Since the edge multiplicity in the projection graph corresponds to the
number of paths of length $2$ in the bipartite graph, we define $E_{k}^{\left(
P2\right)  }:=\mathbb{E}\#\left\{  \text{ }\left(  P,P^{\prime}\right)
:\text{ there are exactly }k\text{ paths of length 2 between }P\text{ and
}P^{\prime}\right\}  $ and $E^{\left(  P2\right)  }:=\sum$ $kE_{k}^{\left(
P\right)  }.$ For fixed $P$ and $P^{\prime}$ with parameters $\mu$ and
$\mu^{\prime}$ the expected number of paths of lenght $2$ between the two
vertices is given by
\begin{equation}
\sum\limits_{\nu}\frac{c^{2}}{N^{2}}\mu\mu^{\prime}\nu^{2}\left\vert G_{\nu
}\right\vert
\end{equation}
and therefore the expected total number of $2$ paths in the $P$-partition is
\begin{align}
E^{\left(  P2\right)  } &  =\sum\limits_{\mu,\mu^{\prime}}\left\vert D_{\mu
}\right\vert \left\vert D_{\mu^{\prime}}\right\vert \sum\limits_{\nu}%
\frac{c^{2}}{N^{2}}\mu\mu^{\prime}\nu^{2}\left\vert G_{\nu}\right\vert \\
&  =\sum\limits_{\mu,\mu^{\prime}}\sum\limits_{\nu}\frac{C_{O}C_{P}^{2}%
M}{C_{op}\left(  \mu\mu^{\prime}\right)  ^{\alpha-1}\nu^{\beta-2}}
\mathperiod
\end{align}
On the other hand, we have for the probability of an edge between $P$ and
$P^{\prime}$ in the P-projection graph the estimate
\begin{align}
\Pr\left\{  P\sim P^{\prime}\right\}   &  =1-\prod\limits_{\nu}\left(
1-\frac{c^{2}}{N^{2}}\mu\mu^{\prime}\nu^{2}\right)  ^{\left\vert G_{\nu
}\right\vert }\\
&  \simeq1-\exp\left(  -\sum_{\nu}\frac{C_{O}c^{2}\mu\mu^{\prime}}{C_{op}%
M\nu^{\beta-2}}\right)
\end{align}
and hence for the expected total number of edges $E$
\begin{equation}
E\simeq\sum\limits_{\mu,\mu^{\prime}}\frac{C_{P}^{2}M^{2}}{\left(  \mu
\mu^{\prime}\right)  ^{\alpha}}\left(  1-\exp\left(  -\sum_{\nu}\frac
{C_{O}c^{2}\mu\mu^{\prime}}{C_{op}M\nu^{\beta-2}}\right)  \right)
\mathperiod
\end{equation}
Several cases are now possible. For $\beta>3$ and $\alpha>2$, it is easy to
see that $\lim\limits_{N\rightarrow\infty}\frac{E^{\left(  P2\right)  }}{E}=1$
and higher edge multiplicities have essentially zero probability. 

The
situation is different if either condition is violated, since in this
case $E^{\left(  P2\right)  }-E$ diverges and can become of the same order as
$E.$ For instance, we obtain for $\beta<3,\alpha<2$
\begin{align}
E^{\left(  P2\right)  }-E &  \simeq\sum\limits_{\mu,\mu^{\prime}}^{\mu_{\max}%
}\frac{C_{P}^{2}M^{2}}{\left(  \mu\mu^{\prime}\right)  ^{\alpha}}\sum_{k\geq
2}\frac{\left(  -1\right)  ^{k}}{k!}\left[  \sum_{\nu}^{\nu_{\max}}\frac
{C_{O}c^{2}\mu\mu^{\prime}}{C_{op}M\nu^{\beta-2}}\right]  ^{k}\\
&  \simeq\sum\limits_{\mu,\mu^{\prime}}^{\mu_{\max}}\frac{const\cdot M^{2}%
}{\left(  \mu\mu^{\prime}\right)  ^{\alpha}}\sum_{k\geq2}\frac{\left(
-1\right)  ^{k}}{k!}\left[  const\cdot\mu\mu^{\prime}M^{\frac{3}{\beta}%
-2}\right]  ^{k}\\
&  \simeq\sum_{k\geq2}const\cdot\frac{\left(  -1\right)  ^{k}}{k!}M^{\frac
{2}{\alpha}+k\left(  \frac{3}{\beta}+\frac{2}{\alpha}-2\right)  }
\end{align}
From the last formula, we see that the expected edge multiplicity
$\frac{E^{\left(  P2\right)  }}{E}-1$ can become positive for proper choices
of $\alpha$ and $\beta.$ 
We show that $\frac{E}{E^{\left(  p2\right)  }}<1$
under the above assumptions. Since
\begin{align}
E^{\left(  P2\right)  } &  =\sum\limits_{\mu,\mu^{\prime}}\sum\limits_{\nu
}\frac{C_{O}C_{P}^{2}M}{C_{op}\left(  \mu\mu^{\prime}\right)  ^{\alpha-1}%
\nu^{\beta-2}}\\
&  \simeq const\cdot M^{\frac{1}{\alpha}2\left(  2-\alpha\right)  +1+\frac
{1}{\beta}\left(  3-\beta\right)  }\\
&  =const\cdot M^{\frac{4}{\alpha}+\frac{3}{\beta}-2}%
\end{align}
and
\begin{equation}
E\simeq\sum_{k\geq1}const\cdot\frac{\left(  -1\right)  ^{k+1}}{k!}M^{\frac
{2}{\alpha}+k\left(  \frac{3}{\beta}+\frac{2}{\alpha}-2\right)  }
\mathcomma
\end{equation}
one gets
\begin{align}
\frac{E}{E^{\left(  P2\right)  }}  & \simeq1-\sum_{k\geq2}const\cdot
\frac{\left(  -1\right)  ^{k}}{k!}M^{\frac{2\left(  k-1\right)  }{\alpha
}+\frac{3\left(  k-1\right)  }{\beta}-2k}\\
& \simeq1-const\cdot M^{-\frac{2}{\alpha}-\frac{3}{\beta}}\left(  M^{\frac
{2}{\alpha}+\frac{3}{\beta}}-1+o\left(  1\right)  \right)  \\
& =1-const+o\left(  1\right)
\mathperiod
\end{align}
Since the involved constant is positive we get the desired result. A more carefully analysis, which will be part of a
forthcoming paper, shows that one also obtains a power law for the edge
multiplicity, as observed in the simulations.

\subsection{Random intersection graphs and the ``Cameo'' principle}

In this section, we give a possible explanation for the appearance of
power laws in the size distribution. In most models of complex networks with
power-law like degree distributions, one assumes a kind of preferential
attachment rule as in the Albert and Barabasi model. This makes little sense in
our framework. Instead we propose a rule called the ``Cameo Principle'' first
formulated in \cite{6}. 

Before giving an interpretation and motivation we
briefly describe the formal setting. Assign to each project a positive
$\varphi$ distributed random variable (r.v.) $\omega$ and to each organization
a positive $\psi-$ distributed r.v. $\mu$ (note that, in contrast to \sxn{sec:molloyreed}, $\varphi$ and $\psi$ are not the size distributions). We
assume $\varphi$ and $\psi$ to be supported on $\left(  1,\infty\right)  $ and
monotone decaying as $\omega$ and $\mu$ tend to infinity. On the bipartite
graph an edge between an organization O and a project P is then formed with
probability
\begin{equation}
p_{o,p}:=\frac{c_{0}}{\psi^{\alpha}\left(  P\right)  }\cdot\frac{1}%
{\sum\limits_{P}\psi^{-\alpha}\left(  P\right)  }+\frac{c_{1}}{\varphi^{\beta
}\left(  O\right)  }\cdot\frac{1}{\sum\limits_{O}\varphi^{-\beta}\left(
O\right)  }\label{a}
\mathcomma
\end{equation}
where $c_{0}$ and $c_{1}$ are positive constants, $\alpha,\beta\in\left(
0,1\right)  $, and all edges are drawn independently of one another. We are
interested in the properties of the corresponding random P and O-graphs for
typical realizations of the $\omega$ and $\mu$ variable. The word typical is
here understood in the sense of the ergodic theorem, namely we assume
$\frac{1}{N}\sum\limits_{O}\varphi^{-\beta}\left(  O\right)  \sim\int
\varphi^{1-\beta}d\varphi=:C_{0}^{-1}$ and $\frac{1}{M}\sum\limits_{P}%
\psi^{-\alpha}\left(  P\right)  \sim\int\psi^{1-\alpha}d\psi=:C_{1}^{-1}$,
where $N$ and $M$ are the cardinalities of the O- and P-partitions
and $\alpha$ and $\beta$ are such that the integral is bounded. The above formula
reduces then to%
\begin{equation}
p_{o,p}:=\frac{c_{0}\cdot C_{0}}{M\psi^{\alpha}\left(  P\right)  }+\frac
{c_{1}\cdot C_{1}}{N\varphi^{\beta}\left(  O\right)  }%
\mathperiod
\end{equation}
The expected conditional size of a vertex is then given by
\begin{equation}
\mathbb{E}\left(  \left\vert P\right\vert \mid\psi\left(  P\right)  \right)
=\frac{Nc_{0}\cdot C_{0}}{C_{1}M\cdot\psi^{\alpha}\left(  P\right)  }%
+c_{1}\label{y}%
\end{equation}
and
\begin{equation}
\mathbb{E}\left(  \left\vert O\right\vert \mid\varphi\left(  O\right)
\right)  =\frac{Mc_{1}\cdot C_{1}}{C_{0}N\cdot\varphi^{\beta}\left(  O\right)
}+c_{0}\label{x}
\mathperiod
\end{equation}

The interpretation behind the special form of edge probability in \eqn{a} is
the following. The $\omega$ and $\mu$ values describe a kind of attractivity
property inherent to projects and organizations. Thinking in terms of a
virtual project formation process the final set of organizations belonging to
a project $P$ can either join the project actively---in which case the $\mu$
value of $P$ is important---or the organization more passively enters the project on
the request of organizations already involved---in which case the
attractivity $\omega$ of the the corresponding organization is important. 
The attractivity
of an organization could, for instance, be related to its reputation, financial
strength, or quality of earlier projects in which the organization was involved.
Extrapolating from human behavior, it is not directly the $\omega$ or $\mu$
value which enters the pairing probability, but rather the relative frequency
of the $\omega$ or $\mu$ values: the rarer a property, the more
attractive it becomes. This is in essence the content of the ``Cameo''
principle. 

The parameters $\alpha$ and $\beta$ can be seen as a kind of
affinity to following the above rule; for $\alpha,\beta\rightarrow0$ the rule is
switched off and we recover a classical Erd\"{o}s-Renyi intersection graph.
In general the values of $\alpha$ and $\beta$ are themselves quenched random
variables with their own---usually unknown---distribution. As shown in
\cite{7}, only the maximal $\alpha$ and $\beta$ values matter
for the resulting degree distribution of the
graphs. We therefore
restrict ourself in the following to constant values. 

Since the conditional
expectation of the size values (\eqns{y} \andeqn{x}) are proportional to
$\varphi^{-\beta}$ and $\psi^{-\alpha}$, we have to estimate their
induced distribution. It can be shown \cite{8} that
$z:=\varphi^{-\beta}\left(  \omega\right)  $ is asymptotically distributed
with density $z^{-\left(  1+\frac{1}{\alpha}+o\left(  1\right)  \right)  }$ 
when $\varphi\left(  \omega\right)  $ decays monotone and faster than any power
law to zero as $\omega\rightarrow\infty$. When $\varphi\left(
\omega\right)  $ is itself a power-law distribution with exponent $\gamma$, the
resulting distribution for $z$ will be $z^{-\left(  1+\frac{1}{\alpha}%
-\frac{1}{\alpha\gamma}+o\left(  1\right)  \right)  }$. Therefore, the induced
distribution is always a power law and independent of the details of $\varphi
$. Applying this result to our model, we obtain immediately a power law
distribution for the size distribution on the P- and O-graphs with
exponents depending essentially only on $\alpha$ and $\beta.$ It is not
difficult to see that, due to the edge independence in the model definition,
the resulting degree distributions are again of power-law type. The Cameo 
\textit{Ansatz} hence generates in a natural way a bipartite graph, where both
projections admit two of the main features of the FP-networks. Furthermore, we
obtain a linear dependence of the mean triangle number
$\triangle_{k}$ on the degree, as in \sxn{sec:randgraphmodelfixedsize}. 

None of the models discussed in section IV can reproduce scale-free
distribution of the edge multiplicity with the same low exponent as observed
in each of the FP networks. It will be interesting to see whether the
inclusion of   memory effects  like the "My friends are your friends"
principle \cite{10} will change the picture. 

\section{Conclusions} \label{sec:conclusions}

In this work, we have described research collaboration networks determined from research projects funded by the European Union. The networks are large in terms of size, complexity, and economic impact. We observed numerous characteristics known from other complex networks, including scale-free degree distribution, small diameter, and high clustering. Using a random intersection-graph model, we were able to reproduce many properties of the actual networks. The empirical and theoretical investigations together shed light on the properties of these complex networks, in particular that the EU-funded R\&D networks match well 
with typical realizations of random graph models characterized by just four parameters: the size, edge number, exponent of project-projection degree distribution, and exponent of organization-projection degree distribution.

In terms of real-world interpretation, the present analysis yields three major insights. First, based on the fact that the size distribution of projects did not 
change significantly between the Framework Programs, any possible changes in project formation rules---which we do not know at this stage---did not affect the aggregate 
structure of the resulting research networks. Second, the fact that integration between collaborating organizations has increased over time, as measured by the average 
clustering coefficient, indicates that Europe has already been moving towards a more 
closely integrated European Research Area in the earlier Framework Programs. Finally, the 
fact that a sizeable number of organizations collaborate more than once in each 
Frame Program shows 
that there appears to be a kind of robust backbone structure in place, which may 
constitute the core of the European Research Area.

In terms of application, the present results suggest a number of extensions. First, it is 
essential to learn more about the properties of the vertices in our networks. To what extent
can they be characterized and classified? What kind of structural patterns emerge if we 
add this information? Second, we need to know more about the micro-structure of 
the networks. In which areas are the networks highly clustered and where does this 
clustering come from? What kind of subgroups can be identified? Third, we need to learn more 
about where the observed distribution of edge multiplicity comes from. Finally, it would be 
desirable to explicitly include edge weights into the analysis. Presumable, actors who 
collaborate more frequently are more proximate to each other than actors who collaborate 
only once. This may significantly impact the structural features we are able to observe, as well 
as the conclusions we might draw concerning the link between network structure and function.

\begin{acknowledgments}
We would like to acknowledge support from the Portuguese Funda\c{c}\~ao para a Ci\^encia e 
a Tecnologia  
(Bolsa de Investiga\c{c}\~ao SFRH/BPD/9417/2002 
and FEDER/POCTI-SFA-1-219), ARC systems research (W4570000294-3),
and from the VW Stiftung (I/80496). We thank Ph.~Blanchard and L.~Streit for useful discussions and commentary. Portions of this work were done at the Vienna Thematic Institute for Complexity and Innovation, EXYSTENCE Network of Excellence: IST-2001-32802. 

\end{acknowledgments}

%: Figures

%\begin{figure}
%	\centering
%	\includegraphics{\figpath orgdegdeg_dep}
%	\caption{}
%	\label{fig:}
%\end{figure}

%\begin{figure}
%	\centering
%	\includegraphics{\figpath orgdegdeg}
%	\caption{}
%	\label{fig:}
%\end{figure}

%
%\begin{figure}
%	\centering
%	\includegraphics{\figpath projdegdeg_dep}
%	\caption{}
%	\label{fig:}
%\end{figure}

%\begin{figure}
%	\centering
%	\includegraphics{\figpath projdegdeg}
%	\caption{}
%	\label{fig:}
%\end{figure}

\end{document}